\documentclass[twocolumn,showpacs,preprintnumbers,amsmath,amssymb]{revtex4}

\usepackage{graphicx}
\usepackage{dcolumn}
\usepackage{bm}

\begin{document}

\def\piminuspiplus{1.025}
\def\piminuspiplusstaterr{0.006}
\def\piminuspiplussyserr{0.018}
\def\kminuskplus{0.95}
\def\kminuskplusstaterr{0.03}
\def\kminuskplussyserr{0.03}
\def\pbarp{0.73}
\def\pbarpstaterr{0.02}
\def\pbarpsyserr{0.03}

\def\npart{317}
\def\nparterr{10}

\def\abcorrgheisha{5.0}
\def\abcorrfluka{2.1}
\def\abcorr{3.5}
\def\aberr{1.5}


\preprint{APS/123-??}

\title{Centrality and pseudorapidity dependence of elliptic flow for charged hadrons in Au+Au collisions at $\sqrt{s_{_{NN}}}=200$~GeV}
\author{	
B.B.Back$^1$,
M.D.Baker$^2$,
M.Ballintijn$^4$,
D.S.Barton$^2$,
R.R.Betts$^6$,
A.A.Bickley$^7$,
R.Bindel$^7$,
A.Budzanowski$^3$,
W.Busza$^4$,
A.Carroll$^2$,
M.P.Decowski$^4$,
E.Garc\'{\i}a$^6$,
N.K.George$^{1,2}$,
K.Gulbrandsen$^4$,
S.Gushue$^2$,
C.Halliwell$^6$,
J.Hamblen$^8$,
G.A.Heintzelman$^2$,
C.Henderson$^4$,
D.J.Hofman$^6$,
R.S.Hollis$^6$,
R.Ho\l y\'{n}ski$^3$,
B.Holzman$^2$,
A.Iordanova$^6$,
E.Johnson$^8$,
J.L.Kane$^4$,
J.Katzy$^{4,6}$,
N.Khan$^8$,
W.Kucewicz$^6$,
P.Kulinich$^4$,
C.M.Kuo$^5$,
W.T.Lin$^5$,
S.Manly$^8$,
D.McLeod$^6$,
A.C.Mignerey$^7$,
M.Nguyen$^2$
R.Nouicer$^6$,
A.Olszewski$^3$,
R.Pak$^2$,
I.C.Park$^8$,
H.Pernegger$^4$,
C.Reed$^4$,
L.P.Remsberg$^2$,
M.Reuter$^6$,
C.Roland$^4$,
G.Roland$^4$,
L.Rosenberg$^4$,
J.Sagerer$^6$,
P.Sarin$^4$,
P.Sawicki$^3$,
W.Skulski$^8$,
P.Steinberg$^2$,
G.S.F.Stephans$^4$,
A.Sukhanov$^2$,
J.-L.Tang$^5$,
M.B.Tonjes$^7$,
A.Trzupek$^3$,
C.M.Vale$^4$,
G.J.van~Nieuwenhuizen$^4$,
R.Verdier$^4$,
G.I.Veres$^4$,
F.L.H.Wolfs$^8$,
B.Wosiek$^3$,
K.Wo\'{z}niak$^3$,
A.H.Wuosmaa$^1$,
B.Wys\l ouch$^4$\\
\vspace{3mm}
\small
$^1$~Argonne National Laboratory, Argonne, IL 60439-4843, USA\\
$^2$~Brookhaven National Laboratory, Upton, NY 11973-5000, USA\\
$^3$~Institute of Nuclear Physics PAN, Krak\'{o}w, Poland\\
$^4$~Massachusetts Institute of Technology, Cambridge, MA 02139-4307, USA\\
$^5$~National Central University, Chung-Li, Taiwan\\
$^6$~University of Illinois at Chicago, Chicago, IL 60607-7059, USA\\
$^7$~University of Maryland, College Park, MD 20742, USA\\
$^8$~University of Rochester, Rochester, NY 14627, USA\\
}

\date{27th October, 2005}
\begin{abstract}\noindent
  This paper describes the measurement of elliptic flow for charged particles in Au+Au collisions at $\sqrt{s_{_{NN}}}=200$~GeV using the PHOBOS detector at the Relativistic Heavy Ion Collider (RHIC). The measured azimuthal anisotropy is presented over a wide range of pseudorapidity for three broad collision centrality classes for the first time at this energy. Two distinct methods of extracting the flow signal were used in order to reduce systematic uncertainties. The elliptic flow falls sharply with increasing $|\eta|$ at 200~GeV for all the centralities studied, as observed for minimum-bias collisions at $\sqrt{s_{_{NN}}}=130$~GeV.
\end{abstract}


\maketitle

It is widely accepted that a very dense and possibly new state of matter is being created in central Au+Au collisions \cite{dAhighpt} at the Relativistic Heavy-Ion Collider (RHIC) at Brookhaven National Laboratory. The azimuthal anisotropy in the distribution of produced particles (``flow'') is a consequence of the initial spatial asymmetry of the collision zone and subsequent rescattering processes which convert this to a final momentum anisotropy.  Measurements of flow are therefore sensitive to the system early in the collision and to its dynamical evolution.

There is an extensive data set on flow results from RHIC 
\cite{Star_v2_vsnpart_vspt_130,Star_v2_Pid_130,Star_v2_KLambda_130,Star_v2_Cummulant_130,Star_v2_AzimuthalCorr_130,Star_v2_pid_200,Star_v1_200,Phenix_v2_twopartcor_130,Phenix_v2_pid_200,phobos130flow,phobosflowlimitfrag}, but as of yet, the least understood result is the pseudorapidity and energy dependence of the elliptic flow, $v_{2} \left( \eta \right)$, measured over an extended $\eta$-range~\cite{phobos130flow,phobosflowlimitfrag}. Recently some theoretical progress has been made by introducing a longitudinal dependence in the source shape and/or by assuming incomplete thermalization away from $\eta=0$ \cite{recent_theory_Heinz,recent_theory_Csanad}. In this paper, we extend our earlier measurements by examining the dependence of the $v_2(\eta)$ shape on the collision centrality. The elliptic flow of charged hadrons has been studied using data from the PHOBOS detector during the 2001 Au+Au run of RHIC.
In addition to the ``hit-based'' method previously used in \cite{phobos130flow}, a new ``track-based'' method was developed and employed to improve the accuracy of the measurement and provide a valuable consistency check of the hit-based analysis. 

The PHOBOS detector consists of silicon pad detectors arranged in single and multiple-layer configurations surrounding the interaction region, as described in~\cite{phobosdetector}. The two multiple-layer magnetic spectrometer tracking arms are configured with a field free region near the interaction vertex followed by tracking inside the magnet. This leads to two classes of found tracks with different acceptances. ``Straight-line'' tracks cover \mbox{$0<\eta<1.8$} with an azimuthal acceptance of $\Delta\phi\approx22^\circ$ centered at $\phi = 0^\circ$ and $180^\circ$. ``Curved'' tracks cover \mbox{$0<\eta<1.5$} with a variable azimuthal acceptance of $\Delta\phi\approx20^\circ$. Details of the tracking procedure are given elsewhere~\cite{phobos200highpt}. The single layer configuration includes the octagonal multiplicity detector (OCT) with \mbox{$|\eta|<3.2$} and six annular silicon ring multiplicity detectors (RINGS), with \mbox{$3.0<|\eta|<5.4$}. The rings and most of the octagon have full azimuthal coverage except near the middle of the detector (mid-rapidity for nominal vertices) where the azimuthal coverage drops by a factor of two.

Two sets of scintillating paddle counters were used for triggering and centrality determination~\cite{phobosprl,phobosPhysRevC,whitepaper}. In addition an online vertex trigger was employed, using two sets of \v{C}erenkov detectors. The hit-based method required events whose collision vertex ($v_{\rm{z}}$) was centered at $-34$~cm away from the nominal vertex position, along the beam axis~\cite{phobos130flow}. The vertex trigger enabled a special sample ($\sim 1$~million triggers) of such events to be taken. The track-based method required events with vertices within about 10~cm of the nominal vertex position, which allowed a large fraction of the 2001 Au+Au data set at 200~GeV ($\sim 25$ million triggers) to be used. The minimum-bias sample for the hit-based method consists of all triggered events that have a valid reconstructed vertex. This engenders biases similar to those discussed in~\cite{phobos130flow} and leads to the average number of participants $\langle N_{part} \rangle$ given in Table~\ref{table1}. For the track-based method, only the fraction of the cross-section unbiased by trigger and vertex inefficiencies is used to form the minimum-biased sample. The average number of participants for this method is also given in Table~\ref{table1}. For the centrality dependent $v_2(\eta)$  analysis, the data samples were subdivided into the three centrality classes given in Table \ref{table1}. The top $3\%$ of the cross-section, where the flow signal is smallest, was omitted to reduce the resulting statistical and systematic errors on the most central bin. Differences in the average number of participants between the two methods, for the same fraction of the Au+Au cross-section, occur because the track-based method is track weighted whereas the hit-based method is event weighted.  This results in slightly higher $\langle N_{part} \rangle$ values for the track-based method, which are insignificant given the systematic error in $\langle N_{part} \rangle$. For both methods the resulting centrality classes are unbiased. The summary of the number of events used is also given in Table \ref{table1}.

\begin{table*}
\begin{tabular}{|c|c|c|c|c|c|c|}
\toprule
Centrality & \multicolumn{3}{c|}{Hit-based} & \multicolumn{3}{c|}{Track-based} \\
& $\% \sigma_{Au+Au}$ & $<N_{part}>$ & Number Events & $\% \sigma_{Au+Au}$ & $<N_{part}>$ & Number  Events \\ \colrule 
minimum-bias & -- & $ 205$ & 34,727 & $0-50$ & $236$ & 5,050,778 \\  
central & $3-15$ & $288$ & 11,221 & $3-15$  &  $294$ & 1,439,923\\ 
mid-central & $15-25$ & $199$ & 7,550 & $15-25$ & $202$ & 1,230,394\\ 
peripheral & $25-50$ & $111$ & 10,127 & $25-50$ & $115$ & 3,087,599\\ 
\botrule
\end{tabular}
\caption{Characteristics of the event samples used in the two flow analyses of 200-GeV Au+Au collisions. The systematic error in $\langle N_{part} \rangle$ is approximately $\pm 4$ participants. \label{table1}}
\end{table*}

Monte Carlo (MC) simulations of the detector performance based on the Hijing \cite{hijing} event generator and GEANT 3.21\cite{GEANT} simulation package were used for systematic error studies.

\begin{figure}
  \centerline{   
      \includegraphics[width=0.5\textwidth]{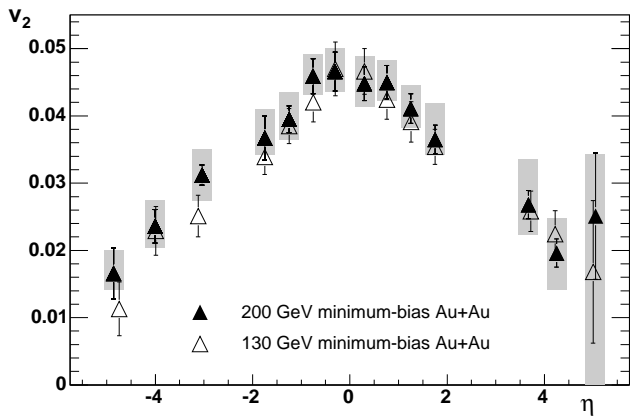}}
  \caption{Elliptic flow as a function of pseudorapidity ($v_{2} (\eta)$) for charged hadrons in minimum-bias collisions at $\sqrt{s_{_{NN}}}=130$~GeV (open triangles) \cite{phobos130flow} and $200$~GeV (closed triangles). One sigma statistical errors are shown as the error bars. Systematic errors (90\% C.L.) are shown as gray boxes only for the 200~GeV data.}
  \label{v2_eta}
\end{figure}

Figure \ref{v2_eta} shows the minimum-bias result for the \mbox{200-GeV} data using the hit-based method (as described in \cite{phobos130flow}). The data show a steady decrease in $v_2$ with increasing $|\eta|$, similar to that seen at the lower energy of $\sqrt{s_{_{NN}}}=130$~GeV (also shown). No significant difference in shape or magnitude is seen within the systematic errors. The ratio of $v_{2}$ at $\sqrt{s_{_{NN}}}=200$~GeV compared to 130~GeV, averaged over all $\eta$, is  $1.04 \pm 0.03 \rm{(stat.)} \pm 0.04 \rm{(syst.)}$. 

The track-based method correlates the azimuthal angle of tracks that traverse the spectrometer, $\phi_{trk}$, with the event plane as measured in the octagon, $\Psi_{2}$, event by event. The method used is based upon the scheme described by Poskanzer and Voloshin \cite{PoskVolo}, where the strength of the flow is given by the $n^{th}$ Fourier coefficient of the particle azimuthal angle distribution
\begin{equation}
  \frac{dN}{d(\phi_{trk}-\Psi_{R})} \sim 1 + \sum_{n} 2v_{n} \cos{\left[ n \left( \phi_{trk}-\Psi_{R} \right) \right]}.
\label{eq1}
\end{equation}
In this analysis only the $n=2$ component is studied and the true reaction plane, $\Psi_{R}$, is approximated by the event plane $\Psi_{2}$.

The use of tracking requires events with vertices near the nominal vertex range \mbox{($-8~{\rm cm}<v_z<10~{\rm cm}$)} to ensure maximum track acceptance in the spectrometer. Only the parts of the OCT detector with complete azimuthal acceptance (i.e. those away from mid-rapidity) are used to determine the reaction plane. Two sub-events, symmetric in $\eta$ and of equal charged particle multiplicity, are used to determine the event plane resolution. The sub-event sizes are vertex dependent, resulting in a resolution correction that is both centrality and vertex dependent. The resulting sub-event ranges lie between \mbox{$2.05<|\eta|<3.2$}, and are widely separated, thus greatly reducing the effects of any short range non-flow correlations. The event plane is determined using
\begin{equation}
  \Psi_{2}= \frac{1}{2} \tan^{-1}{\left( \frac{\sum_i w_i \sin(2\phi_{i})}{\sum_i w_i \cos(2\phi_{i})} \right)},
\label{eq2}
\end{equation}
where $\phi_{i}$ is the $i^{th}$ hit's measured angle, and the sums run over all hits in both sub-events. The sub-events are combined for the event plane determination in order to maximize its resolution. Vertex dependent corrections, some determined on an event-by-event basis, are used as weights ($w_{i}$)~\cite{phobos130flow} in order to remove acceptance and occupancy biases. The resulting distributions of event plane angles are found to be flat within $2\%$.

To determine the $v_2$ coefficient, the measured $\frac{dN}{d(\phi_{track}-\Psi_{2})}$ distribution is divided by a mixed event distribution in order to remove detector related effects, such as non-uniformities in the azimuthal acceptance of the spectrometer:
\begin{equation}
\left. \frac{dN}{d \Delta \phi} \right|_{\rm{measured}} \Big/ \left. \frac{dN}{d \Delta \phi} \right|_{\rm{mix}} \sim 1+ 2 \left( \frac{v_{2}}{C_{\rm{res}}} \right) \cos(2 \Delta \phi), \label{eqnew}
\end{equation} 
where $\Delta \phi$ denotes $\phi_{track}-\Psi_{2}$ and $C_{\rm{res}}$ is the event plane resolution correction. The $\left. \frac{dN}{d \Delta \phi} \right|_{\rm{mix}}$ distribution, with zero flow, is constructed using an event mixing technique, where the $\phi_{track}$ of tracks in one event are subtracted from the $\Psi_{2}$ of another event.

Normalized $\Delta \phi$ distributions,\mbox{$\left. \frac{dN}{d \Delta \phi} \right|_{\rm{measured}} \Big/ \left. \frac{dN}{d \Delta \phi} \right|_{\rm{mix}}$ ,} and $C_{\rm{res}}$, are determined as a function of vertex position and for fine centrality bins ($\sim 5\%$ of the cross-section per bin) since the event mixing technique requires similarity of the class of events examined. The centrality bin and vertex dependent event plane resolution correction $C_{\rm{res}}(\mbox{centrality},v_{z})$ are determined using the sub-event technique \cite{PoskVolo} as
\begin{equation}
  C_{\rm{res}}(\mbox{centrality},v_{z}) = 
 \frac{1}{\sqrt{2} \alpha \sqrt{\left< \cos{\left[ 2 \left( \psi^{\eta<0}_{2}-\psi^{\eta>0}_{2}\right) \right]} \right>}},
\label{eq3}
\end{equation}
where $\psi^{\eta<0}_{2}$ and $\psi^{\eta>0}_{2}$ are the event planes from each sub-event. The $\alpha\sqrt{2}$ factor converts the single sub-event resolution correction into a combined sub-event resolution correction. The $\alpha$ factor is sometimes approximated as unity, but this approximation can break down, particularly when the event resolution is good. Its exact form is given in Refs. \cite{PoskVolo} and \cite{cmv}. For the resolutions measured in this data set, \mbox{$0.95<\alpha<1$}.

After averaging Eq.~\ref{eqnew} over vertex positions and centralities falling into each broad centrality class defined in Table \ref{table1}, the $v_{2}$ coefficient is extracted from the fit to an even-harmonic series. (Including orders higher than $n=2$ in the fit did not effect the extracted $v_{2}$.)
It should be noted that the resulting $v_2$ is a track weighted result over the broad centrality classes, since limited statistics precluded the $v_2$ from being determined for each fine centrality bin and then event weighted. For further details on this technique see \cite{cmv}.

Extensive MC simulations have shown that the magnitude and shape of the flow signal are correctly reproduced by this method. No further corrections to the measured $v_2$ coefficient are necessary, such as potential corrections due to the density of particles or suppression corrections due to backgrounds, as required in the hit-based method. 

\begin{figure}[t]
  \centerline{
      \includegraphics[width=0.5\textwidth]{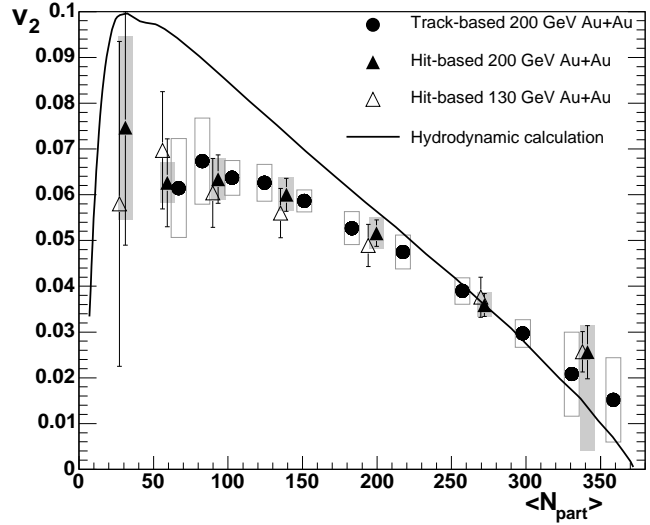}
  }
  \caption{ Elliptic flow ($v_2(\left| \eta \right|<1)$) as a function of $\langle N_{part} \rangle$ determined by the track-based method (closed circles) and hit-based method (closed triangles) for Au+Au collisions at $200$~GeV. The open triangles are the results from Au+Au collisions at 130~GeV. One sigma statistical errors are shown as the error bars (within the points for the track-based method); gray and open boxes show systematic uncertainties (90\% C.L.) for the 200-GeV results from the hit-based and track-based methods, respectively.  The line shows a calculation from hydrodynamics \cite{HydroLineModel} at $\sqrt{s_{_{NN}}}=200$~GeV.
  }
  \label{fig_v2_npart}
\end{figure}

In addition to the sources of systematic errors considered for the hit-based analysis~\cite{phobos130flow}, other studies performed for the track-based method include analysis of the effects related to tracking, such as varying cuts on the distance of closest approach of tracks to the collision vertex, differences between results obtained from the two spectrometer arms, momentum resolution, and dependence on the bending direction. Additionally, contributions due to the vertex dependency of the resolution corrections, different beam orbit conditions, and errors of the fit parameters were also accounted for. 

Figure \ref{fig_v2_npart} shows the centrality dependence of the $v_{2}$ determined using the straight line track-based method over a range of \mbox{$0<\eta<1$}, allowing a direct comparison with the same result using the hit-based technique. The two techniques agree well over the full range of centrality. The curve in Figure \ref{fig_v2_npart} shows a hydrodynamic calculation for Au+Au collisions at $\sqrt{s_{_{NN}}}=200$~GeV~\cite{HydroLineModel}. As seen for Au+Au collisions at 130~GeV~\cite{Star_v2_vsnpart_vspt_130} (open triangles), the 200-GeV results at mid-rapidity are consistent with expectations from hydrodynamic models. There is no significant difference between the 130- and 200-GeV data in either the shape or magnitude of $v_2$ at mid-rapidity as a function of centrality within the errors.

\begin{figure}
\centerline{
\includegraphics[width=0.5\textwidth]{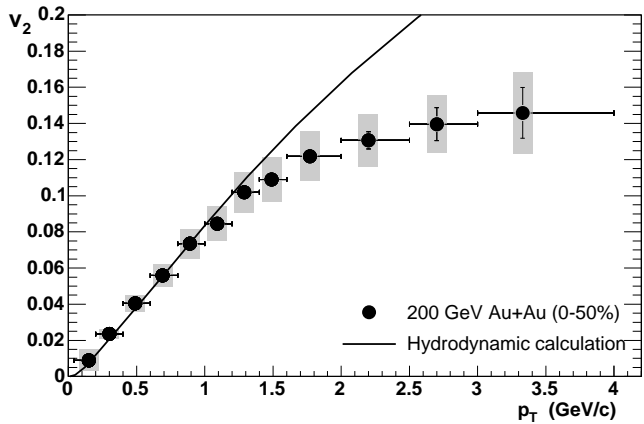}}
\caption{Elliptic flow as a function of transverse momentum ($v_{2}(p_{T})$) for charged hadrons with \mbox{$0<\eta<1.5$}  for the most central $50\%$ of the 200~GeV Au+Au inelastic cross section. The one sigma statistical errors are shown as the error bars. The gray boxes represent the systematic errors (90\% C.L.).  The data points are located at the average $p_{T}$ position within a $p_{T}$ bin whose size is given by the horizontal error bars. The curve shows a calculation from hydrodynamics \cite{HydroLineModel}.}
\label{fig_v2_pt}
\end{figure}

Using tracks that traverse the full field region of the spectrometer, the transverse momentum dependence of the flow strength $v_{2}(p_{T})$ can be measured. This is shown in Figure~\ref{fig_v2_pt} for the top $0-50\%$ centrality for tracks averaged over the range \mbox{$0<\eta<1.5$}. The curve shows the prediction of a hydrodynamical model \cite{HydroLineModel}. As previously observed \cite{Star_v2_vsnpart_vspt_130}, the $v_2$ rises as $p_{T}$ increases and at $p_{T}$ above $1.5$~GeV/c  tends to flatten out well below the hydrodynamic curve. 

In these analyses, the reaction plane is determined in sub-events that are at different pseudorapidities from those where the $v_2$ is measured. This should significantly reduce the contribution of any non-flow effects to the measured $v_2$, particularly those due to short-range correlations. Comparisons of the $v_{2}(p_{T})$ result to the reaction plane and cumulant methods results from reference \cite{Star_v2_Cummulant_130}, averaged over a similar centrality range, show that our result is most consistent with the one obtained with the four particle cumulant method \cite{QM2004Flow}, suggesting that our track-based methodology is indeed largely immune to non-flow effects over the range \mbox{$|\eta|<1.5$}.

\begin{figure}
  \centerline{
    \includegraphics[width=0.5\textwidth]{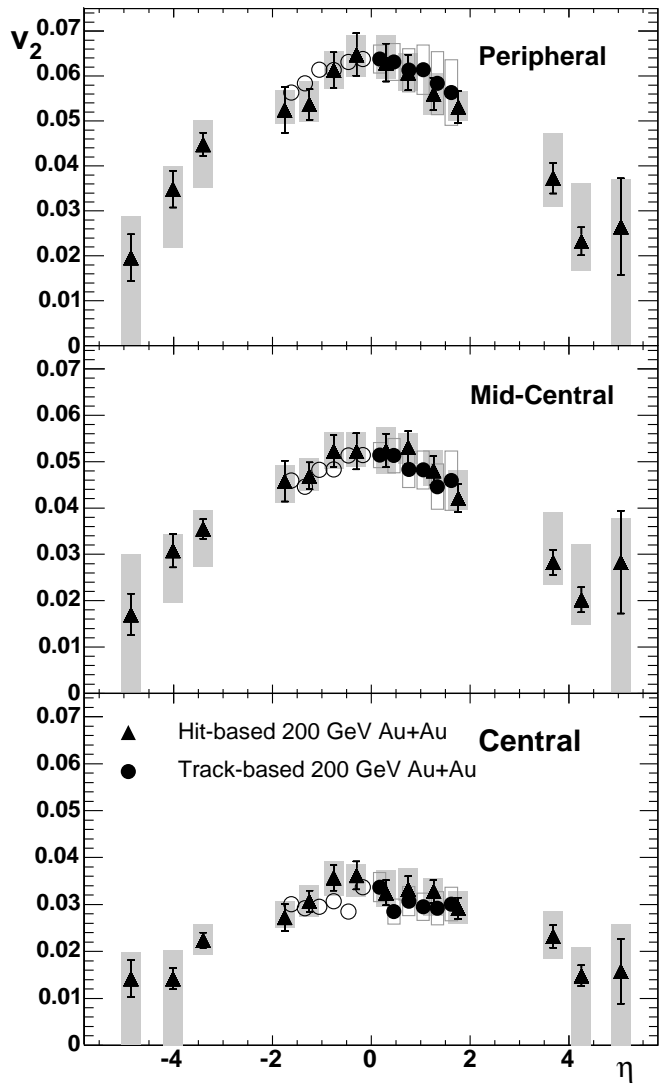}}
  \caption{Elliptic flow as a function of pseudorapidity ($v_{2}(\eta)$) for charged hadrons from 200-GeV Au+Au collisions for the three different centrality classes described in the text, ranging from peripheral to central ($25-50\%$, $15-25\%$, $3-15\%$) from top to bottom. The triangles are the results from the hit-based method, and the circles are from the track-based method. The open circles are the track-based results reflected about mid-rapidity. One sigma statistical errors are shown as the error bars (within the points for the track-based method); the gray and open boxes show the systematic uncertainties (90\% C.L.) for the hit-based and track-based methods, respectively.}
\label{fig_v2_eta_both}
\end{figure}

Figure \ref{fig_v2_eta_both} shows $v_2(\eta)$ for three centrality classes as defined in Table \ref{table1}. Excellent agreement is seen across all of the centrality classes over the range of overlap suggesting that our hit-based method is also minimally affected by non-flow effects around mid-rapidity.

\begin{figure}
  \centerline{
      \includegraphics[width=0.5\textwidth]{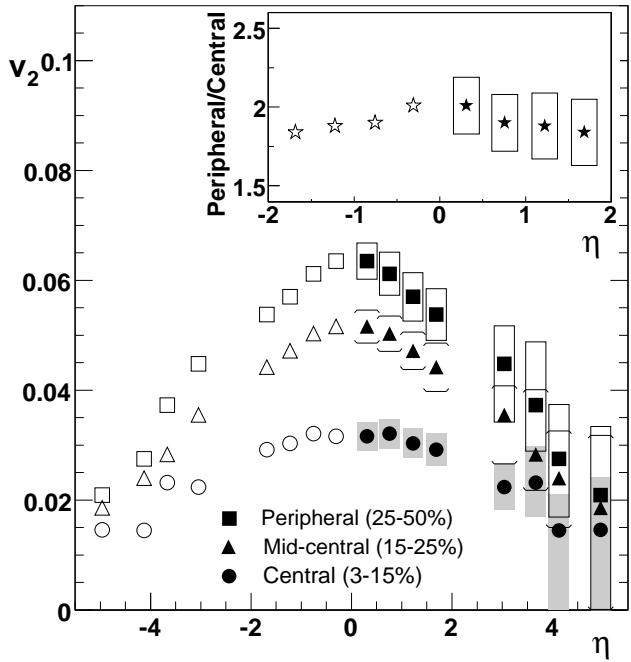}}
  \caption{Elliptic flow as a function of pseudorapidity ($v_{2}(\eta)$) from 200-GeV Au+Au collisions for the three centrality bins ($3-15\%$ circles, $15-25\%$ triangles, $25-50\%$ squares). Data for \mbox{$\eta>0$} are determined by reflecting the hit-based results about mid-rapidity and then combining them with the track-based results and are shown with the corresponding combined 90\% C.L. statistical and systematic uncertainties. The same data are reflected around $\eta = 0$ and shown as open symbols.  In the range where the methods overlap, the insert shows the ratio of the peripheral to central results, with the appropriate 90\% C.L. combined uncertainties. }
  \label{fig_v2_eta_comb}
\end{figure}

To examine how the shape of the distribution changes with centrality, the results of the hit-based method and track-based methods are combined. Although obtained in the same experiment, the measurements should effectively be considered independent of each other due to the very different methods and elements of the PHOBOS detector used; hence the results for each method are combined with the reasonable assumption that the errors are uncorrelated. First, the hit-based results that are approximately an equal $\eta$ distance away from mid-rapidity are combined (e.g. $\eta = -4.87$ with $\eta = +5.06)$, weighted by their statistical uncertainties. The points at $\langle \eta \rangle=-3.05$ and $\langle \eta \rangle=+3.67$ are just reflected due to the lack of symmetry of these points around $\eta=0$. The track-based results at $\eta = +0.17$ and $\eta = +0.46$ are also combined to give $v_{2}$ at $\eta = +0.31$, and similarly for $\eta = +1.05$ and $\eta = +1.34$, to give a $v_{2}$ at $\eta = +1.20$. The hit-based results and the track-based results with similar $\eta$ binning are averaged, weighted by their combined statistical and systematic uncertainties. The resulting data are shown in Figure \ref{fig_v2_eta_comb}. The pseudorapidity dependence of $v_{2}$ for the 3 centrality bins is similar to that observed in Fig.\ref{v2_eta} for minimum-bias data. For peripheral collisions, $v_{2}$ clearly already has a non-zero slope over the range \mbox{$-2<\eta<2$}. The overall shape of $v_{2}(\eta)$ is not strongly centrality dependent within the uncertainties, appearing to differ only by a scale factor. This is illustrated in the insert of Fig.~\ref{fig_v2_eta_comb}, which shows that the ratio of the peripheral to central data around mid-rapidity is approximately constant.  However, it should be noted that the central data around mid-rapidity is also consistent with a flat distribution, given the uncertainties.


In summary, we have measured the centrality dependence of $v_{2}(\eta)$ in Au+Au collisions at $\sqrt{s_{_{NN}}}=200$~GeV. Excellent agreement with the track-based method further validates the use of the hit-based method. This method allowed for the  study of the $v_{2}(\eta)$ dependence over the large range of $\eta$ covered by the PHOBOS single-layer silicon detectors. The 200-GeV results clearly show that $v_{2}$ decreases with increasing $\left| \eta \right|$, as seen for the 130-GeV Au+Au collisions. From comparisons of the $v_{2}(p_{T})$ results with four particle cumulant results we conclude that our flow measurements are largely immune to non-flow effects, over the range \mbox{$|\eta|<1.5$.}

The predominant features of the $v_{2}(\eta)$ distribution do not change significantly as a function of centrality from  $\langle N_{part}\rangle$  $\sim 290$ to $\langle N_{part}\rangle$  $\sim 110$. The flow still falls off as one moves away from mid-rapidity. It is hoped that this data can be used to more fully understand the strong $\eta$ dependence of the $v_{2}$ flow component.

This work was partially supported by U.S. DOE grants DE-AC02-98CH10886,
DE-FG02-93ER40802, DE-FC02-94ER40818, DE-FG02-94ER40865, DE-FG02-99ER41099, 
and W-31-109-ENG-38, US NSF grants 9603486, 9722606 and 0072204, Polish KBN 
grant 1-P03B-06227, and NSC of Taiwan contract NSC 89-2112-M-008-024.

\end{document}